\documentclass[fleqn,usenatbib]{mnras}
\usepackage{amsmath}
\usepackage{xspace}
\usepackage{mathptmx}
\usepackage{txfonts}
\usepackage{xcolor}

\usepackage[T1]{fontenc}
\usepackage{ae,aecompl}

\usepackage{graphicx}	
\usepackage{amssymb}	
\usepackage{booktabs}
\usepackage{wasysym}

\let\oldhref\href
\renewcommand{\href}[2]{\oldhref{#1}{\hbox{#2}}}

\definecolor{colorl1}{RGB}{0, 51, 153}
\definecolor{colorl2}{RGB}{153, 0, 0}
\definecolor{colorl3}{RGB}{179, 179, 0}
\definecolor{colorl4}{RGB}{51, 102, 0}

\definecolor{colorw1}{RGB}{51, 102, 255}
\definecolor{colorw2}{RGB}{255, 51, 0}
\definecolor{colorw3}{RGB}{255, 214, 51}
\definecolor{colorw4}{RGB}{51, 204, 51}

\newcommand{\hMpc}{{\ifmmode{h^{-1}{\rm Mpc}}\else{$h^{-1}$Mpc}\fi}}
\newcommand{\Mpc}{{\ifmmode{{\rm Mpc}}\else{Mpc}\fi}}
\newcommand{\hkpc}{{\ifmmode{h^{-1}{\rm kpc}}\else{$h^{-1}$kpc}\fi}}
\newcommand{\kpc}{{\ifmmode{ {\rm kpc}}\else{{\rm kpc}}\fi}}
\newcommand{\kms}{{\ifmmode{ {\rm km\,s^{-1}}}\else{ ${\rm km\,s^{-1}}$}\fi}}
\newcommand{\hMsun}{{\ifmmode{h^{-1}{\rm {M_{\astrosun}}}}\else{$h^{-1}{\rm{M_{\astrosun}}}$}\fi}}
\newcommand{\Msun}{{\ifmmode{{\rm M}_{\astrosun}}\else{${\rm M}_{\astrosun}$}\fi}}
                        
\newcommand{\Mhalo}{{\ifmmode{M_{\rm halo}}\else{$M_{\rm halo}$}\fi}}
\newcommand{\Rvir}{{\ifmmode{R_{\rm vir}}\else{$R_{\rm vir}$}\fi}}
\newcommand{\Mvir}{{\ifmmode{M_{\rm vir}}\else{$M_{\rm vir}$}\fi}}
\newcommand{\Mstar}{{\ifmmode{M_{\rm star}}\else{$M_{\rm star}$}\fi}}
\newcommand{\Vrot}{{\ifmmode{V_{\rm rot}}\else{$V_{\rm rot}$}\fi}}
\newcommand{\ltsima}{$\; \buildrel < \over \sim \;$}
\newcommand{\gtsima}{$\; \buildrel > \over \sim \;$}
\newcommand{\lsim}{\lower.5ex\hbox{\ltsima}}
\newcommand{\gsim}{\lower.5ex\hbox{\gtsima}}

\def\lesssim{\mathrel{\hbox{\rlap{\hbox{\lower4pt\hbox{$\sim$}}}\hbox{$<$}}}}
\def\gtrsim{\mathrel{\hbox{\rlap{\hbox{\lower4pt\hbox{$\sim$}}}\hbox{$>$}}}}

\newcommand{\beq}{\begin{equation}}
\newcommand{\eeq}{\end{equation}}
\def\beqa{\begin{eqnarray}}
\def\eeqa{\end{eqnarray}}
\def\LCDM{\ensuremath{\Lambda}CDM}

\def\head{ \vbox to 0pt{\vss \hbox to 0pt{\hskip 440pt\rm
      LA-UR-10-07069\hss} \vskip 25pt}}

\def \kms {\ifmmode  \,\rm km\,s^{-1}\else $\,\rm km\,s^{-1}$\fi }
\def \kpc {\ifmmode  {\,\rm kpc}  \else ${\rm  kpc}$ \fi  }  
\def \hkpc {\ifmmode  {h^{-1}\rm kpc}  \else ${h^{-1}\rm kpc}$ \fi  }  
\def \hMpc {\ifmmode  {h^{-1}\rm Mpc}  \else ${h^{-1}\rm Mpc}$ \fi  }  
\def \Mpch {\ifmmode  {h^{-1}\rm Mpc}  \else ${h^{-1}\rm Mpc}$ \fi  }  
\def \Msun {\ifmmode {\rm M}_{\astrosun} \else ${\rm M}_{\astrosun}$ \fi} 
\def \hMsun {\ifmmode h^{-1}\,\rm M_{\astrosun} \else $h^{-1}\,\rm M_{\astrosun}$ \fi}
\def \Gyr {\ifmmode\, \rm Gyr \else $\,$Gyr \fi}

\def \LCDM {\ifmmode \Lambda{\rm CDM} \else $\Lambda{\rm CDM}$ \fi}
\def \sig8 {\ifmmode \sigma_8 \else $\sigma_8$ \fi} 
\def \OmegaM {\ifmmode \Omega_{\rm m} \else $\Omega_{\rm m}$ \fi} 
\def \Omegab {\ifmmode \Omega_{\rm b} \else $\Omega_{\rm b}$ \fi} 
\def \OmegaL {\ifmmode \Omega_{\rm \Lambda} \else $\Omega_{\rm \Lambda}$\fi} 
\def \Deltavir {\ifmmode \Delta_{\rm vir} \else $\Delta_{\rm vir}$ \fi}
\def \rhocrit {\ifmmode \rho_{\rm crit} \else $\rho_{\rm crit}$ \fi}
\def \rhou {\ifmmode \rho_{\rm u} \else $\rho_{\rm u}$ \fi}
\def \zc {\ifmmode z_{\rm c} \else $z_{\rm c}$ \fi}

\def\Mstar {\ensuremath {M_{*}(<r_{23.5})}~}
\def\r23_5 {\ensuremath {r_{23.5}}~}

\title[Galaxies lacking Dark Matter] {Creating a galaxy lacking dark matter in a  dark matter dominated universe}

\author[A.V. Macci\`o et al.]{Andrea V. Macci\`o$^{1,2,3}$\thanks{E-mail: maccio@nyu.edu},
Daniel Huterer Prats$^{1}$, Keri L. Dixon$^{1,2}$ , Tobias Buck$^4$, 
\newauthor Stefan Waterval$^{1,2}$, Nikhil Arora$^5$, St\'ephane Courteau$^5$, Xi Kang$^{6}$\\
$^{1}$New York University Abu Dhabi, PO Box 129188, Abu Dhabi, United Arab Emirates \\
$^2$Center for Astro, Particle and Planetary Physics (CAP$^3$), New York University Abu Dhabi\\
$^3$Max-Planck-Institut f\"ur Astronomie, K\"onigstuhl 17, 69117 Heidelberg, Germany \\
$^4$Leibniz-Institut f\"ur Astrophysik Potsdam (AIP), An der Sternwarte 16, D-14482 Potsdam, Germany\\
$^5$Department of Physics, Engineering Physics and Astronomy, Queen's University, Kingston, ON K7L 3N6, Canada\\
$^6$Zhejiang University-Purple Mountain Observatory Joint Research Center for Astronomy, Zhejiang University, Hangzhou, 310027, China
}

\date{Accepted XXX. Received YYY; in original form ZZZ}

\pubyear{2020}

\begin{document}

\label{firstpage}
\pagerange{\pageref{firstpage}--\pageref{lastpage}}
\maketitle
\begin{abstract}

We use hydrodynamical cosmological simulations to show that it is possible to create, via tidal interactions, galaxies lacking dark matter in a dark matter dominated universe. 
We select dwarf galaxies from the NIHAO project, obtained in the standard Cold Dark Matter model and use them as initial conditions for simulations of satellite-central interactions. 
After just one pericentric passage on an orbit with a strong radial component, NIHAO dwarf galaxies can lose up to 80 per~cent of their dark matter content, but, most interestingly, their central ($\approx 8$~kpc) dark matter to stellar ratio changes from a value of ${\sim}25$, as expected from numerical simulations and abundance matching techniques, to roughly unity as reported for NGC1052-DF2 and NGC1054-DF4. 
The stellar velocity dispersion drops from ${\sim}30$~\kms\ before infall to values as low as $6\pm 2$~\kms. These, and the half light radius around 3~kpc, are in good agreement with observations from van Dokkum and collaborators. Our study shows that it is possible to create a galaxy {\it without} dark matter starting from typical dwarf galaxies formed in a dark matter dominated universe, provided they live in a dense environment. 

\end{abstract}

\begin{keywords}
cosmology: theory -- dark matter -- galaxies: formation -- galaxies: kinematics and dynamics -- methods: numerical
\end{keywords}

\section{Introduction}\label{sec:introduction}

A galaxy can be defined as a concentration of gas and stars sitting at the bottom of a dark matter potential well \citep[e.g.][]{Mo2010}.
In the current cosmological model, Dark Matter (DM) outnumbers baryons (gas and stars) by a factor of roughly one to five, and, at the zeroth order, galaxies are expected to reflect this mass ratio. 
Other effects also contribute to the depletion of baryons in a collapsed object, either by preventing gas from being accreted e.g the UV background \citep{Bullock2001}, or by expelling gas from the galaxy via supernova (SN)
explosions \citep[e.g.][]{Tollet2019} and AGN feedback \citep[e.g.][]{Vogelsberger2014}.
As a result, galaxies are all mostly DM-dominated\footnote{This statement holds for galaxies in their outskirts where the enclosed mass is indeed dominated by a non-baryonic component. Closer to the center, galaxies display a wide range of mass fractions ranging from baryon to dark matter dominated \citep{Courteau2015}.}.

Such a prediction has been challenged by the discovery of two galaxies (NGC 1052-DF2 and NGC 1052-DF4: DF2 and DF4 hereafter) that seem to {\it lack} dark matter \citep{VanDokkum2018, VanDokkum2019} in their central regions.
These galaxies have a stellar mass of $\sim 10^8$ \Msun and are satellites of NGC 1052 \citep[but see][for a different interpretation for DF4]{Montes2020}, a massive ($M_* \approx 10^{11}$ \Msun) elliptical galaxy \citep{Muller2019}.

Given their stellar mass, these objects are predicted to form 
in a dark matter halo of a mass ${\sim}5 \times 10^{10}$ \Msun \citep{Moster2013};
this mass translates into an expected (stellar) velocity dispersion of about 25~\kms. 
Quite surprisingly, using the line-of-sight velocity of $\sim9$ globular clusters up to a distance of 10 kpc, van Dokkum and collaborators have estimated a maximum velocity dispersion below 10~\kms.
When translated back into a mass, such a low velocity dispersion suggests a dark matter to stellar mass ratio of two to three, quite far from values of $\sim$30-40 at that radius 
expected from galaxy formation theory for a halo velocity dispersion of 25~\kms \citep[e.g.][]{Behroozi2013}.
Note that the values of 400 for the dark matter to stellar mass ratio reported in the abstract of \citet{VanDokkum2018} is expected at the virial radius ($\approx 50-60$ kpc); therefore, it is somewhat misleading to compare this value with results obtained on smaller scales ($\approx 10$~kpc).

Several ideas have been advanced to explain these DM-deficient galaxies. 
For example, high speed collisions of gas rich dwarf galaxies \citep{Silk2019, Shin2020} or self-interacting dark matter \citep{Yang2020} might be possible channels for the formation of such objects. Since both observed galaxies are confirmed satellites of the same massive elliptical, a tidal origin may be the most plausible explanation. 

Large numerical simulations have been extensively searched for the existence of dark matter deficient galaxies, with somewhat inconclusive results: 
\citet{Jing2019} used the EAGLE and Illustris simulations to find the $\sim$2 percent of the satellites in the mass range $10^9< M_*< 10^{10}$~\Msun to be DM-poor.
However, resolution limitations  \citep[for example, \citealt{Jing2019} did not use the
higher resolution EAGLE simulations as in][]{Ploeckinger2018} restricted the use of a stellar mass
to nearly a magnitude larger than the observed one (i.e. $10^9$ versus $10^8~\Msun$). 
A similar exercise was carried out by \citet{Haslbauer2019} with the Illustris suite who found a much lower occurrence of DM-deficient galaxies, around 0.1 percent in the full simulated volume. 

In order to overcome the limited resolution provided by large-volume simulations (at the scale of interest), \citet{Ogiya2018} performed a series of
controlled $N$-body simulations of the tidal interaction between NGC 1052 and a smaller satellite galaxy aimed at representing the progenitor of DF2 and DF4.
Under specific orbit conditions, these authors showed that it is indeed
possible to preferentially remove dark matter from the satellite and hence create a DM-poor galaxy.
While promising, this study used ad~hoc assumptions about the DM and stellar profile of the satellites, with a spherical model for all the satellites in setting up the initial conditions for their numerical experiments. Cosmological numerical simulations, on the contrary, predict triaxial haloes on these scales \citep[e.g.,][]{Butsky2016}.

Towards a similar aim, we wish to combine state-of-the-art cosmological numerical simulations of galaxy formation from the NIHAO project \citep[Numerical Investigations of Hundred Astrophysical Objects][]{Wang2015} with hydrodynamical simulations of tidal interactions between a satellite galaxy and the central object.
We shall use as initial conditions for the interaction simulations the output of fully cosmological hydrodynamical simulations of dwarf galaxy formation, and set them on radial orbits around
a central potential aimed at reproducing the properties of NGC 1052.
Besides the orbital trajectory of the satellite, no other free parameters are present in our simulations.

This paper is organized as follows: in Section~\ref{sec:simulations}, we describe the NIHAO sample and the set up for the central-satellite interaction simulations. In Section~\ref{sec:obs}, we revisit the observational data for DF2 and DF4. Our results are then 
presented in Section~\ref{sec:results}, while Section~\ref{sec:conclusions} is devoted
to summarizing and discussing our findings.

\section{Simulations} \label{sec:simulations}

We have selected our galaxies from the NIHAO (Numerical Investigation of Hundred Astrophysical Objects) sample, currently the the largest suite of high-resolution cosmological zoom-in simulations
\citep{Wang2015}.

These simulations have been performed with the {\sc \small gasoline2} code
\citep[described in][]{Wadsley2017} and include metal cooling, chemical enrichment, star formation and feedback from SN and massive stars (the so-called Early Stellar Feedback). 
The cosmological parameters have been set according to 
 \citet{Planck2014}, and each galaxy is resolved with roughly
one million elements (dark matter, gas, and stars) within the virial radius.

NIHAO simulations can successfully reproduce a large variety of galaxy properties like
the stellar-to-halo mass relation \citep{Wang2015}, the disc gas mass and disc size relation \citep{Maccio2016},  the Tully-Fisher relation \citep{Dutton2017}, the diversity of galaxy rotation curves \citep{Santos-Santos2018}, and the the satellite mass function of the Milky Way and M31
\citep{Buck2019}.

Using realistic cosmological hydrodynamical simulations as a starting point 
is one of the advantages of our work with respect to previous attempts
to reproduce galaxies lacking dark matter. High-resolution cosmological simulations naturally provide realistic orbits for both the dark matter and the star particles. Moreover, they do not impose any (unrealistic) assumptions, such as spherical symmetry or dynamical equilibrium for the halo or the stellar disc. 
For example, galaxies with stellar masses around $10^8~$ \Msun are seldom perfectly rotating discs, and they usually show a spheroidal disc component with substantial velocity dispersion \citep{Wheeler2017,Frings2017}.

We have selected three NIHAO galaxies, namely g5.05e10, g6.91e10, and g7.12e10 
(hereafter G1, G2 and G3)\footnote{In the NIHAO project, the name of a galaxy 
reflects its total mass in the low resolution run.}, as initial conditions;
these cover the stellar mass and size (half-light radius) range of DF2 and DF4. 
These simulations have all used the same resolution, with a dark matter and star mass resolution of $m_{\rm dm}=5.1 \times 10^4$ \Msun and $m_*=2.5 \times 10^3$ \Msun and a gravitational softening of $\epsilon_{\rm dm} =310$~pc and $\epsilon_{*} = 132$~pc.
The main structural properties of the three NIHAO galaxies are listed in Table \ref{Table1}.
They have been extracted from the cosmological simulations at $z=0$ and cut at 2.5 times their virial radius \citep{Maccio2019}.

\begin{table}
    \begin{tabular}{ccccccc}
    Name & NIHAO  & $M_*$ & $M_{\rm DM}$ & $r_{\rm h}$ & $\sigma_* (<8$ kpc) \\
        & name & [\Msun] & [\Msun] & [kpc] & [\kms] \\
\hline
\hline
        G1 & g5.05e10 & $9.5 \times 10^7$ & $4.3 \times 10^{10}$ & 1.6 & 27.5 \\
      \hline
  G2 & g6.91e10 & $2.5 \times   10^8$  & $7.1 \times 10^{10}$ & 2.1 & 26.8 \\
\hline
  G3 & g7.12e10 & $6.8 \times 10^8$  & $8.4 \times 10^{10}$ & 2.8 & 37.2 \\
\hline
\end{tabular}
    \caption{Initial NIHAO galaxy properties before infall. $r_{\rm h}$ is the half-light radius. The stellar and dark matter mass are measured within the virial radius of each galaxy.
    }
    \label{Table1}
\end{table}

\begin{figure}
\includegraphics[width=0.47\textwidth]{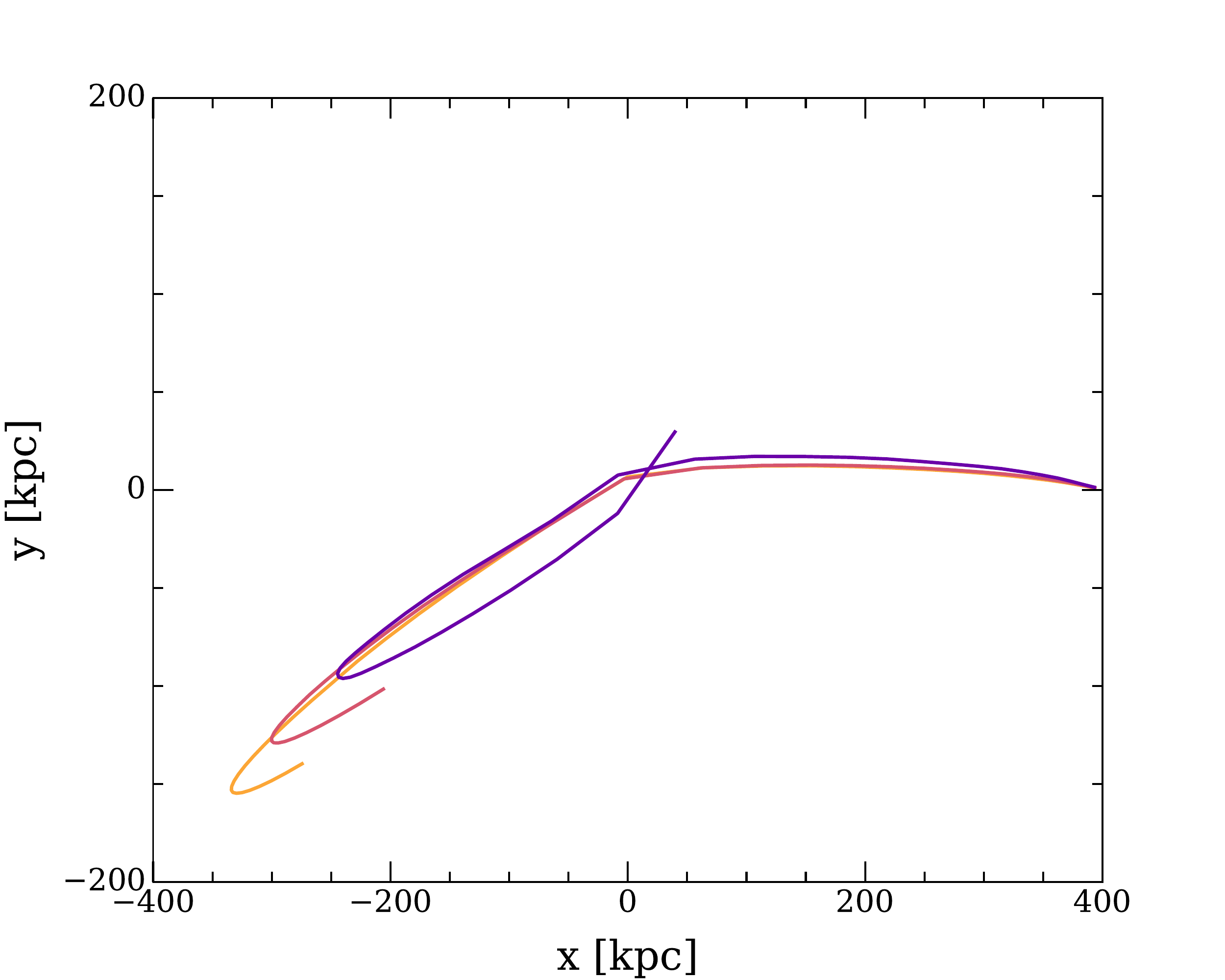}
\vspace{-.25cm}
\caption{Trajectory in the $xy-$plane for orbits O1, O2, and O3 (yellow, red and purple, respectively).
Orbit parameters are listed in Table~\ref{tab:orbits}.
All orbits started at the virial radius of the host halo (400 kpc) and were evolved for 4.5 Gyr
(see text for mores details).}
\label{fig:orbit}
\end{figure}

Following the approach of \citet{Ogiya2018}, we have set those galaxies on radial orbits around a central potential. 
All simulations start at a distance of 400~kpc from the center, which represents the virial radius of the host halo. 
Each orbit is defined by the initial velocity of the galaxy with respect to the virial velocity 
of the host halo ($V_{\rm vir}=230$ \kms); all orbits are in the $xy-$plane and hence have $v_{\rm z}=0$.
A complete list of the orbital parameters is presented in Table \ref{tab:orbits}; each orbit is evolved for 4.5 Gyr to mimic a typical infall time of the satellites around $z\sim1$ \citep[][]{Maccio2010c}. Fig.~\ref{fig:orbit} shows three of the orbits (O1,O2 and O3) in the $xy-$plane that will be analyzed in detail in this paper.

\begin{table}
\centering
\begin{tabular}{cccccc}
        Name &   $v_{\rm x}$ & $v_{\rm y}$ & $v_{\rm z}$ & d [kpc] & symbol \\
\hline
\hline
         O1 &   -0.35 & 0.05 & 0.0 & 7 & $\triangledown$ \\ 
         O2 &   -0.25 & 0.05 & 0.0 & 7 &$ \hexagon$\\ 
         O3 &   -0.25 & 0.07 & 0.0 & 11 & $\varhexagon$ \\ 
         O4 &   -0.4 & 0.04 & 0.0 & 31 & $ \pentagon$\\ 
         O5 &   -0.3 & 0.07 & 0.0 & 29 & $\triangle$ \\ 
         O6 &   -0.35 & 0.04 & 0.0 & 10 & $\Square$ \\ 
         O7 &   -0.2 & 0.07 & 0.0 & 32 & $\Circle$ \\ 
         O8 &   -0.4 & 0.05 & 0.0 & 31 & $\Diamond$ \\
\hline
\hline
\end{tabular}
    \caption{List of a sub-sample of the orbits analyzed in this paper. The orbits are in the $xy-$plane. The values of $v_{\rm x}$ and $v_{\rm y}$ are in units of the virial velocity of the host halo ($V_{\rm vir}=230$~\kms). $d$ is the pericenter distance (for orbit $O3$ 
    the only one with two passages, we report the first passage). The different symbols
    are used to identify the orbits in Fig.~\ref{fig:allorbits}.}
    \label{tab:orbits}
\end{table}


The central potential is fully analytic and composed of two components:
i) a dark matter halo with total mass of $6\times 10^{12}$ \Msun \citep{Forbes2019} and concentration parameter of 8 \citep{Dutton2014} and ii) a Hernquist bulge to describe the stellar component (aka the central massive elliptical galaxy NGC1052), with a total mass of $7.8 \times 10^{10}~\Msun$ and a radius of 0.7~kpc \citep{Muller2019}.

All simulations were performed with the {\sc \small gasoline2} code \citep{Wadsley2017} with the same set up used previously in \citet{Frings2017}, which includes ram pressure, star formation, and stellar+SN feedback \cite[see][for more details]{Frings2017}. 
Due to their low mass, the satellite galaxies are quite gas poor and the inclusion of star formation and related feedback processes is of marginal importance for the results presented in this work. 

\section{Observational Data}
\label{sec:obs}

The observations used in this work are based on the discovery papers of DF2 \citep{VanDokkum2018} and DF4 \citep{VanDokkum2019}.
For each observed galaxy, we use both $\sigma_*$ measurements presented in these papers: the observed velocity dispersion (8.4~\kms\ for DF2 and 5.8~\kms\ for DF4) to which we assign a Poissonian error based on the number of globular clusters used to compute it, and the intrinsic velocity dispersion (3.2~\kms\ for DF2 and 4.2~\kms\ for DF4) and relative errors, as computed by the authors taking into account observational biases \citep[see][for more details]{VanDokkum2018,VanDokkum2019}.

For their two galaxies, DF2 and DF4, \cite{VanDokkum2018,VanDokkum2019} adopted the stellar mass-to-light ratio $M_*/L_V=(2.0\pm0.5){\rm M_{\odot}/L_{\odot}}$ based on the assumption that the galaxies' average stellar population would resemble that of a globular cluster (GC) with a metallicity [$Z$/H]=-1 and age of 11~Gyr. 
It was then confirmed by placing a stellar population at $D=20$~Mpc  that  matches  the  global  properties of NGC1052–DF2 with the adopted $M_*/L_V$.

On the other hand, for typical globular clusters, a metallicity [$Z$/H]=-1 corresponds to [Fe/H] $\sim-1.3\pm0.2$ \citep{Thomas2011}. 
Such a metallicity is quite high for dwarf galaxies \citep{Kirby2015}. 
Therefore, extrapolations involving GC SEDs may not be representative as the stellar populations of dwarf galaxies and GCs (measured through their SEDs) can differ significantly.

One may also obtain typical stellar population parameters of DF2 and DF4 by comparing with other dwarf galaxies.  
To arrive at a representative value of $M_*/L_V$, we first transform the colour reported in \cite{VanDokkum2018} for DF2, namely $V_{606}-I_{814}=0.37\pm0.05$, into the Johnson-Cousin\footnote{\url{https://colortool.stsci.edu/uvis-filter-transformations}} colour of $V-I=0.52\pm0.05$ (this error does not account for the small transformation scatter). This colour is converted into a stellar mass-to-light colour via transformations calibrated for Local Group dwarf galaxies \citep{Z17}.  
The resulting stellar $M_*/L_V$ for DF2 is $M_*/L_V=0.33\pm0.05$.  
While this approach yields a smaller $M_*/L_V$, the methodology does not rely on a globular cluster template.  

In order to consider all possible ranges of stellar population in dwarf galaxies, we also compare with a large database of $M_*/L_V$ estimates for low-mass galaxies. 
A sample of $\sim$400 low-mass galaxies ($\log_{10}(M_*/{\rm M_{\odot}})\leq 9.0$) from the MaNGA survey \citep{MaNGA} yields an average $M_*/L_V=1.1\pm0.2$ (Arora et al., in prep.).  
The colour-mass-to-light models in \cite{Z17} were also used to transform $g-r$ colours of MaNGA galaxies into $M_*/L_V$.  The average, error, and models used in this calculation account for a full range of evolutionary histories of dwarf galaxies.

In order to obtain stellar masses for DF2 and DF4, we have adopted the value $M_*/L_V=1.1\pm0.2$. We have also assumed a systematic error of 0.3 dex for the estimated stellar masses \citep{Roediger2015}. Besides the uncertainty on $M_*/L_V$, this systematic error accounts for the various assumptions involved in transforming light into mass \citep[e.g. IMF, star formation histories, stellar population models, etc.;][]{Courteau2014}.

\section{Results}\label{sec:results}

The key observable in the DF2 and DF4 galaxies is the velocity dispersion of the stars that is then used to determine the total dynamical mass of the object and hence the luminous-to-dark mass ratio.

\begin{figure}
\includegraphics[width=0.47\textwidth]{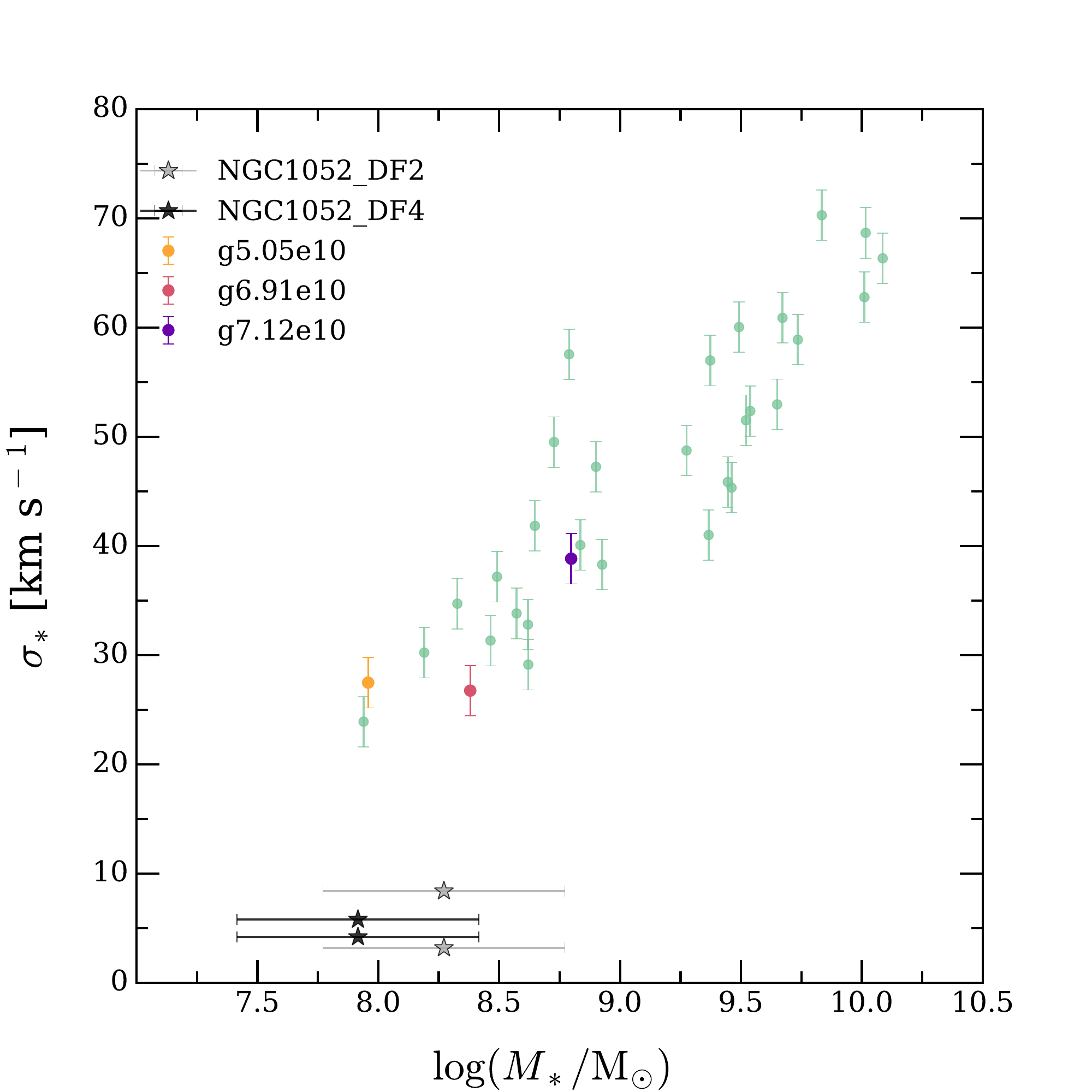}
\vspace{-.25cm}
\caption{The relation between $M_*$ and stellar velocity dispersion computed within 8~kpc (for consistency with the measurements of van Dokkum and collaborators) for NIHAO galaxies. 
Grey and black stars represent the observed values for DF2 and DF4, respectively (see section \ref{sec:obs} for more details).
The three NIHAO galaxies listed in Table~\ref{Table1} are marked with different symbols and colours. 
The velocity dispersion of simulated galaxies is computed along the $x-$axis of the simulation.}
\label{fig:nihao}
\end{figure}

Initially, we searched the NIHAO database for galaxies naturally lacking DM.
Fig.~\ref{fig:nihao} shows the stellar velocity dispersion within 8~kpc vs stellar mass ($M_*$)
for NIHAO galaxies; note that the value of 8~kpc for computing both quantities was chosen for consistency with observational data.
None of the simulated galaxies
have a velocity dispersion comparable to those of DF2 or DF4
in the same  $M_*$ range ($\sim 3\times 10^8$ \Msun);
the predicted velocity dispersion from the NIHAO simulations 
is close to 20-25~\kms, while the observed values are below 10~\kms. 

Fig.~\ref{fig:nihao} suggests that some additional external effects
are needed to bring simulations and observations into agreement, and since NIHAO galaxies have all been selected to be centrals and fairly isolated \citep[see][for more details]{Wang2015}, 
it is natural to study the effect of the environment.
Given that DF2 and DF4 are both satellites of NGC1052, we want to study the effects of tidal interactions in details by performing high resolution central/satellite simulations.

\begin{figure}
\includegraphics[width=0.47\textwidth]{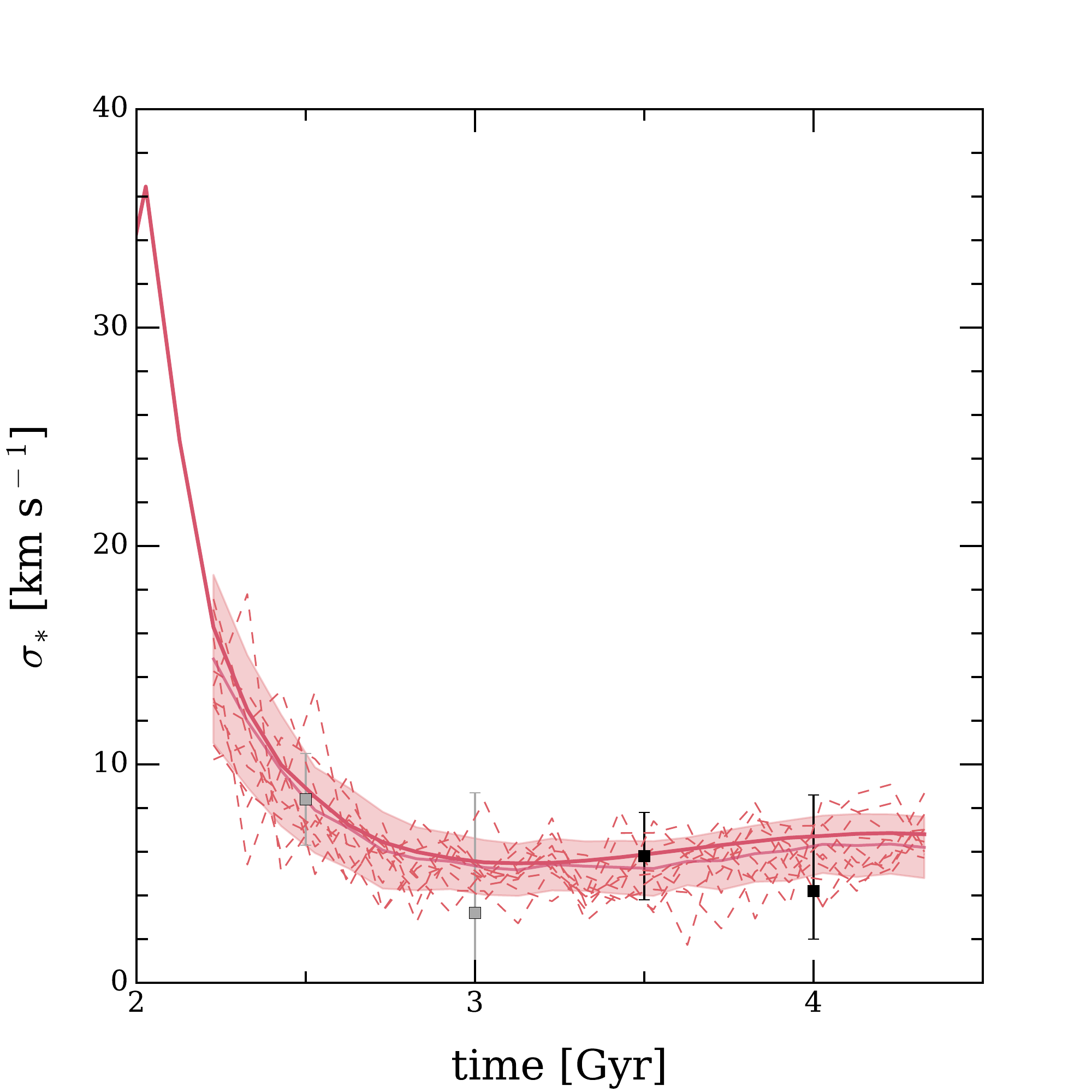}
\vspace{-.25cm}
\caption{Time evolution of the stellar velocity dispersion ($\sigma_*$) within 8 kpc for G2 on the O2 orbit. For clarity, only the last two Gyr after the pericenter passage are shown, 
see Fig.~\ref{fig:sigma} for the complete evolution. The solid red line represents the value for $\sigma_*$ computed using all stars within 8 kpc. The dashed lines show ten different realizations of the velocity dispersion calculations with 9 star particles; the solid darker red line indicates the average and scatter (shaded area) obtained from one hundred realizations of the 9 stars sampling. The grey and black squares show the observational results for DF2 and DF4, respectively, which have been assigned arbitrary infall times from 2.5 to 4 Gyr to increase the readability of the plot (see text for more details).}
\label{fig:sigma1}
\end{figure}

Fig.~\ref{fig:sigma1} shows the evolution of the stellar velocity dispersion ($\sigma_*$) within 8~kpc from the center (consistent with the measurements of van Dokkum and collaborators) 
for G2 on the O2 orbit\footnote{All velocity dispersions shown in the paper are 1D and computed along the $x-$axis of the simulation.}.
The solid line represents $\sigma_*$ computed using all stars within the selected radius, and agrees well with the observational points from \citet{VanDokkum2018,VanDokkum2019}.
On the other hand, the observed data are based on the measurements of the line of sight velocity 
of ten and seven globular clusters for DF2 and DF4, respectively; in order to match observations closely, we have also tried to select nine old and metal-poor stellar particles from our simulations, ensuring they have the same radial distribution as the globular cluster in DF2. The dashed line in Fig.~\ref{fig:sigma1} shows the velocity dispersion obtained this way, which is in good agreement with the mean velocity dispersion (from all stars). There is a scatter of about 2~\kms quite independent of time (the scatter in the figure is based on 100 realizations of the nine point sampling, ten of those realizations are shown in the plot). We will use this scatter due to the limited number of velocity 
tracers in the following as the default error on the dispersion velocity.

Despite being similar parameters, not all galaxy-orbit pairs are equally effective in reducing the dark matter fraction and/or the stellar velocity dispersion. Fig.~4 shows the change in $\sigma_*$ 
as a function of time for three different orbit-galaxy pairs.
Two of them have final values for the stellar velocity dispersion below 10~\kms, while the galaxy on the O3 orbit, despite having a pericenter distance of few kpc, has an almost constant $\sigma_*$ throughout the whole orbit, around 30~\kms.

\begin{figure}
\includegraphics[width=0.47\textwidth]{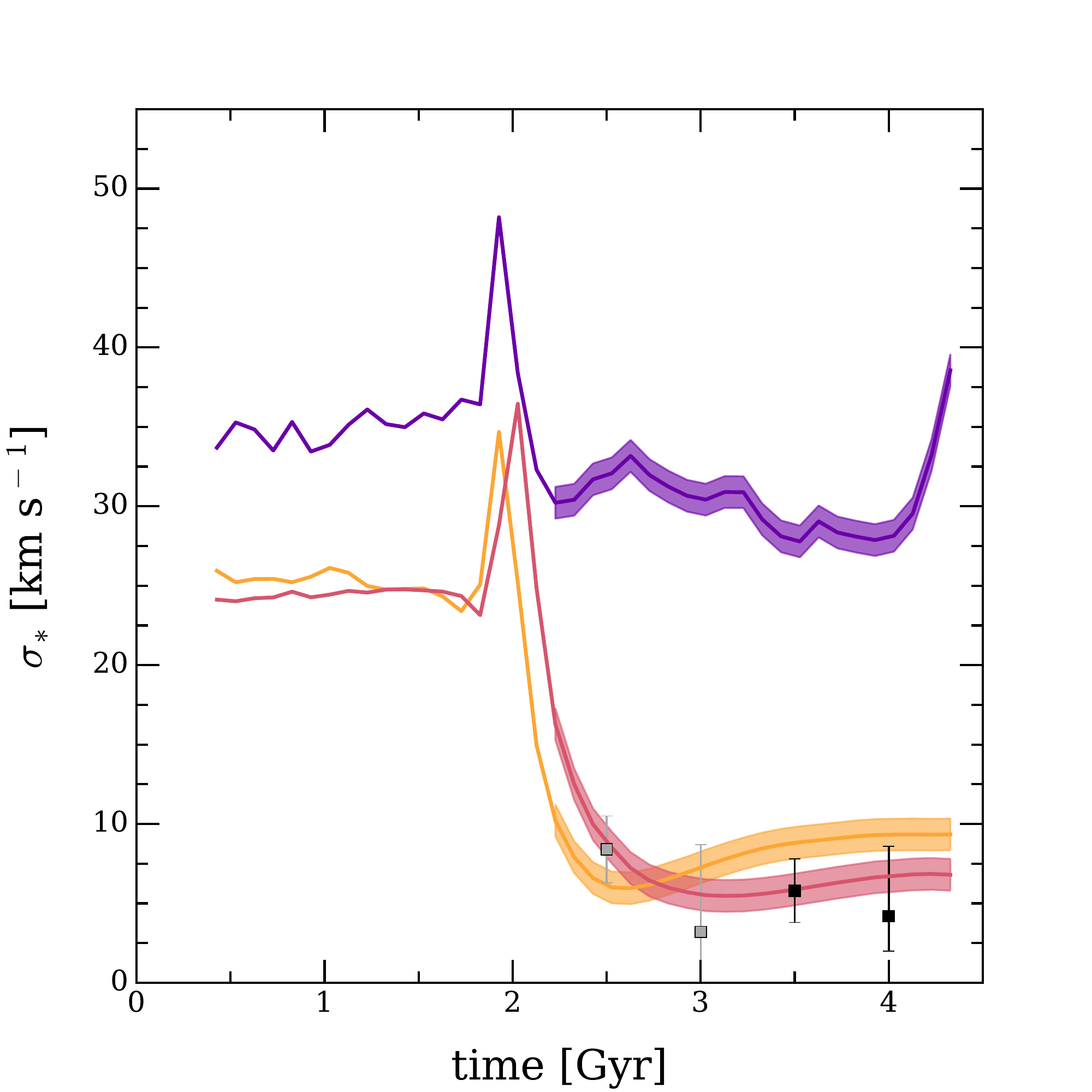}
\vspace{-.25cm}
\caption{Velocity dispersion vs time for G1-O1 (yellow), G2-O2 (red), and 
G3-O3 (purple). The shaded area around each curve represents a constant scatter of 2~\kms due to the limited sampling with N=9 stellar particles, the grey black squares with error bars show the observational results for DF2 and DF4, respectively(see text for more details).
The second spike for G3-O3 is due to the second pericenter passage of the galaxy 
(see Fig.~\ref{fig:orbit}).}
\label{fig:sigma}
\end{figure}

\begin{figure}
\includegraphics[width=0.47\textwidth]{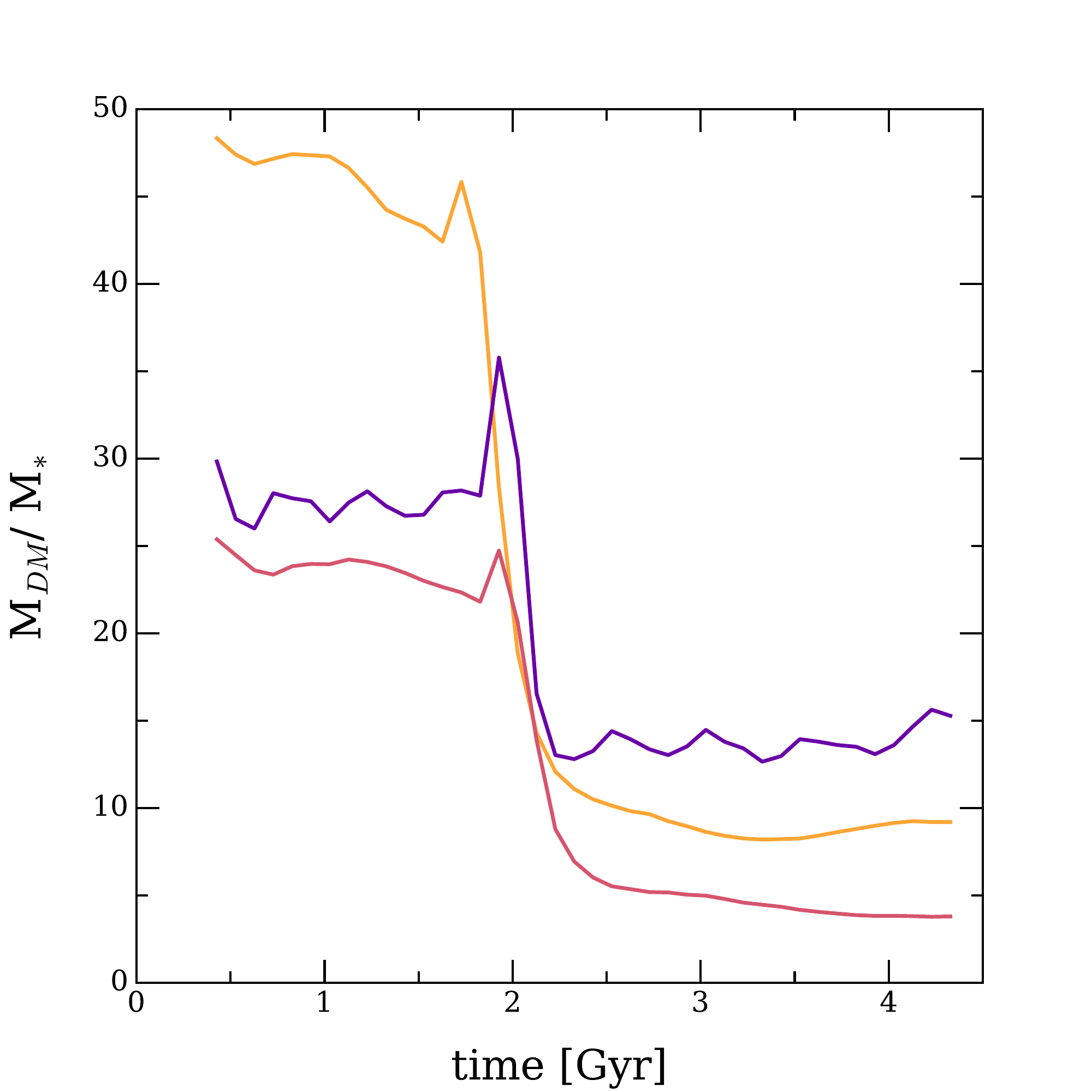}
\vspace{-.25cm}
\caption{Evolution of the dark-to-stellar mass ratio within 8~kpc
for G1-O1 (yellow), G2-O2 (red), and G3-O3(purple).}
\label{fig:mratio}
\end{figure}

The change in the stellar velocity dispersion is related to a substantial change in the mass content in the inner regions of the galaxy. 
Fig.~\ref{fig:mratio} shows the change in the dark-to-stellar mass ratio within a radius of 8~kpc over time.

\begin{figure}
\includegraphics[width=0.47\textwidth]{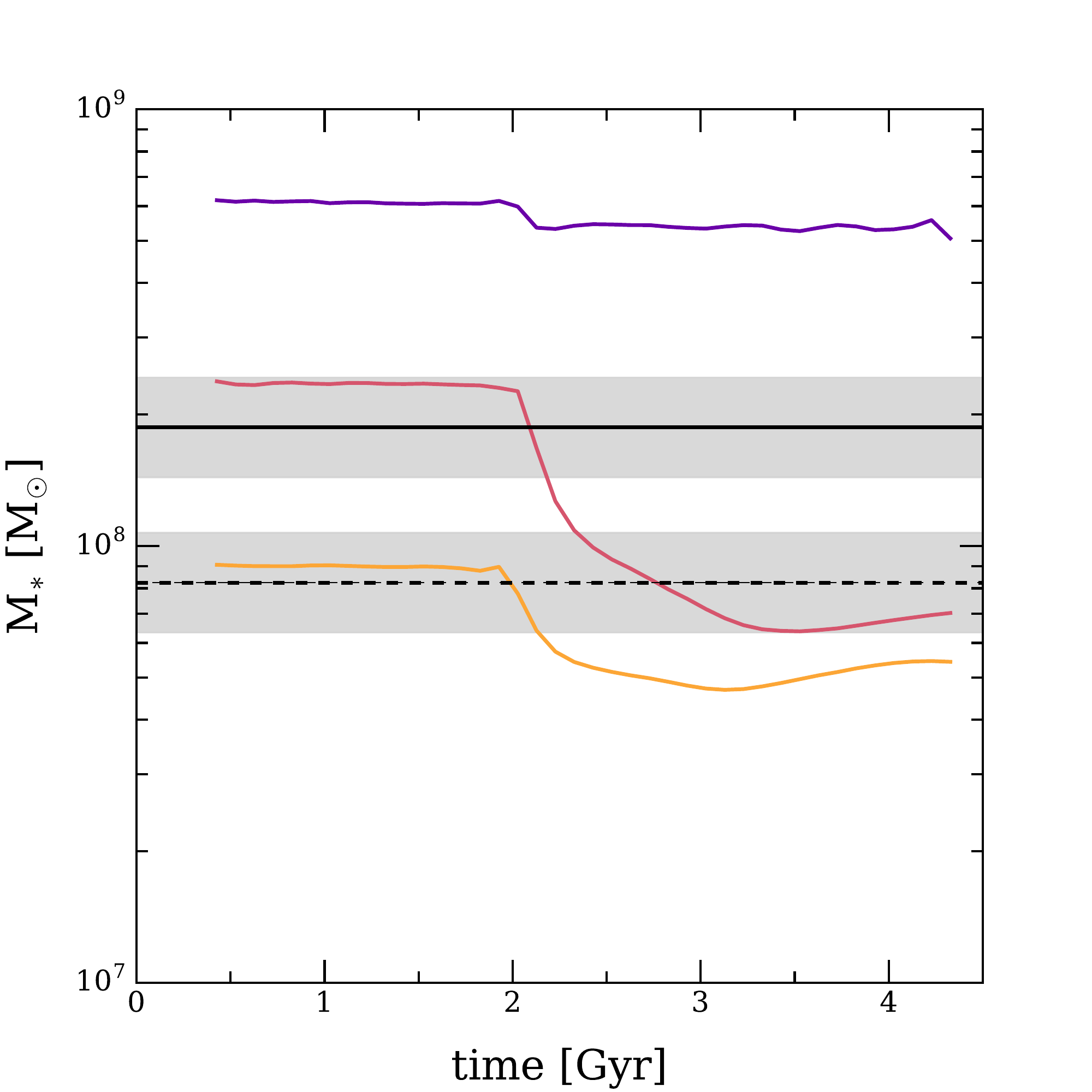}
\vspace{-.35cm}
\caption{Evolution of $M_*$ for the three galaxies. The color coding is the same as in previous figures, the solid and dashed grey lines represent the estimated $M_*$ values for DF2 and DF4, respectively.}
\label{fig:mstar}
\end{figure}

The galaxy orbit combinations with the strongest 
$\sigma_*$ variation are the ones showing the largest reduction 
of the dark-to-stellar mass ratio, which drops
by a factor larger than five in all cases. 
While the dark matter content is reduced by more than a factor of ten, 
the $M_*$ is much more resilient to tidal forces, as shown in
Fig.~\ref{fig:mstar}. The stellar content of all galaxies is 
within a factor of few (on average $\approx 2$) with respect the initial one, 
showing a strong difference in the reaction of the inner ($R<8$~kpc)
distribution of DM and stars to tidal effects.

This difference in the response of the inner stellar component when compared
with the inner dark matter one is related to the very different orbits that DM and stars have within the galaxy.
Dark matter particles are on quite extended orbits \citep[e.g.][]{Jesseit2004}, and particles that are in the inner region of the galaxy at a given time can move quite farther out in the halo at a later time and hence be subject to tidal stripping effects. On the contrary, stars are confined to the inner region and then shielded from tidal perturbations by the dark matter \citep[e.g.][]{Penarrubia2010,Buck2019}. This difference in the orbit distribution 
between stars and DM, is the key ingredient to {\it preferentially remove} dark matter instead of stars from the central region of a galaxy and then create an object that is almost dark matter free. 

\begin{figure}
\includegraphics[width=0.47\textwidth]{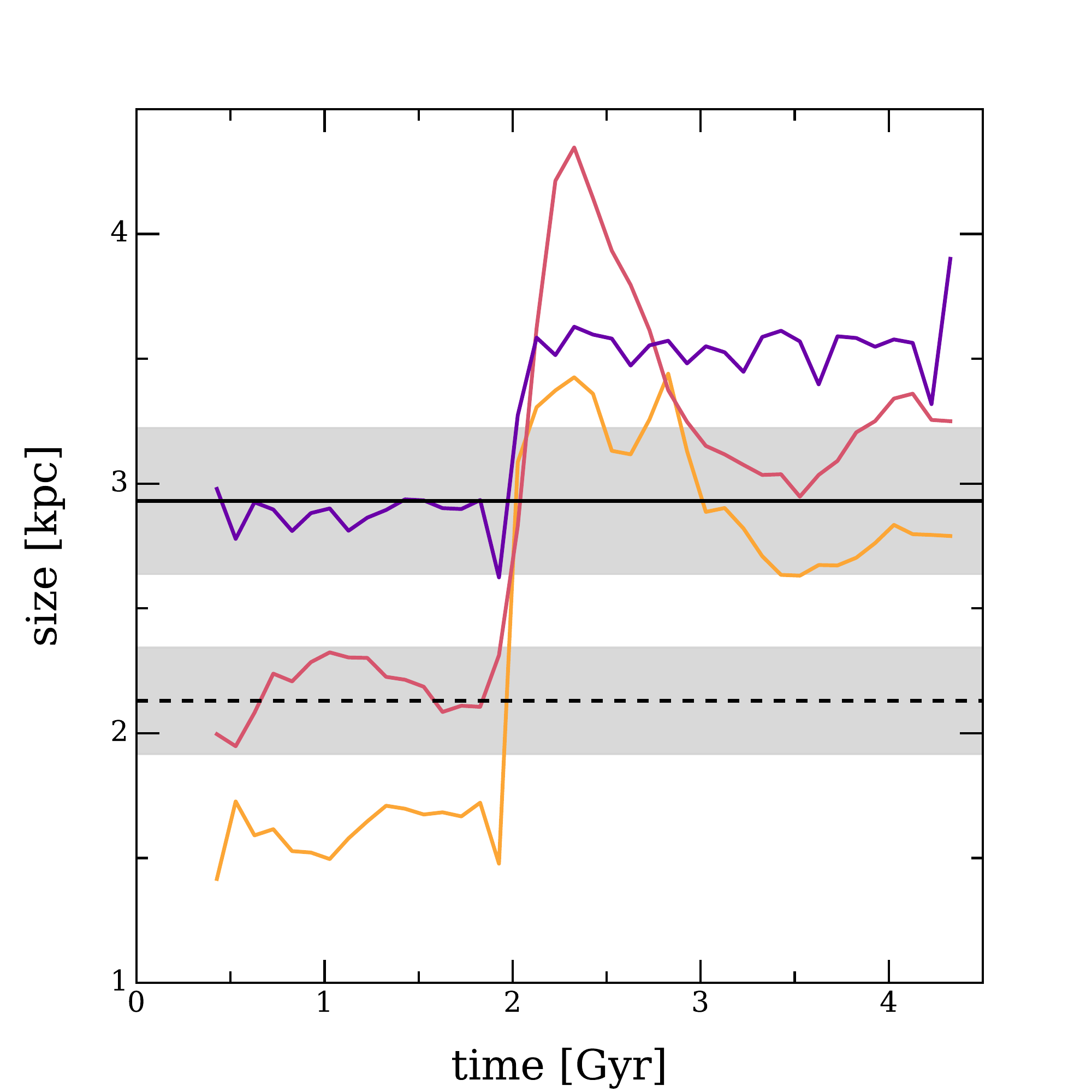}
\vspace{-.25cm}
\caption{Evolution of the half-light radius with time. Simulations 
are color coded as in previous plots. The grey solid and dashed lines represent the measured half-light radius (with errors) for DF2 and DF4, respectively.}
\label{fig:size}
\end{figure}

Another effect of tidal forces is to induce dynamical relaxation in the perturbed object \citep{Chang2013}, which usually leads to an increase in radius.
Fig.~\ref{fig:size} shows the behaviour of the stellar radius as a function of time and compare it to the observational values for DF2 and DF4.

In order to estimate the half-light radius, we first calculate the $V$-band surface brightness profile for each galaxy in face-on alignment. The luminosity of every star particle, given its age, metallicity, and initial mass function, is assigned using the Padova simple stellar population model\footnote{http://stev.oapd.inaf.it/cgi-bin/cmd} \citep{Marigo2008, Girardi2010}. We exclude any stars that are beyond the radial surface brightness threshold of 31~mag~arcsec$^{-2}$, consistent with the DRAGONFLY survey \citep{2016ApJ...830...62M}. We then find the 2D radius within which half the $V$-band light resides. 
Finally, we do not account for dust extinction, which would generally serve to reduce the simulated galaxy radii. It is worth noticing that the exact value of the surface brightness threshold does not strongly affect the estimated radius. 

Our results show quite a large scatter in the response of G1, G2, and G3  to interaction with the central potential.
It is interesting to see if this difference is due to the orbital parameters or to the intrinsic properties of the galaxies.

\begin{figure}
\includegraphics[width=0.47\textwidth]{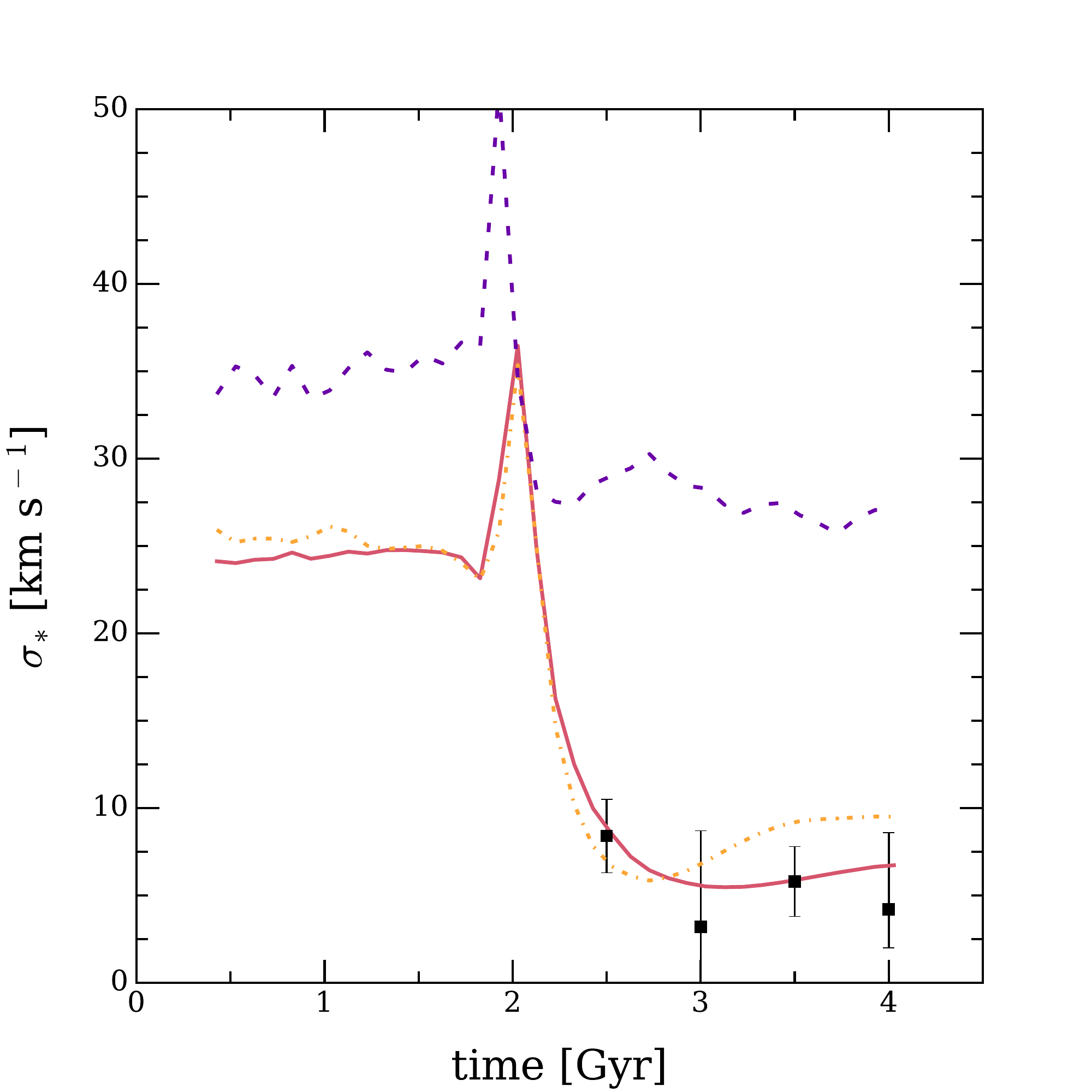}
\includegraphics[width=0.47\textwidth]{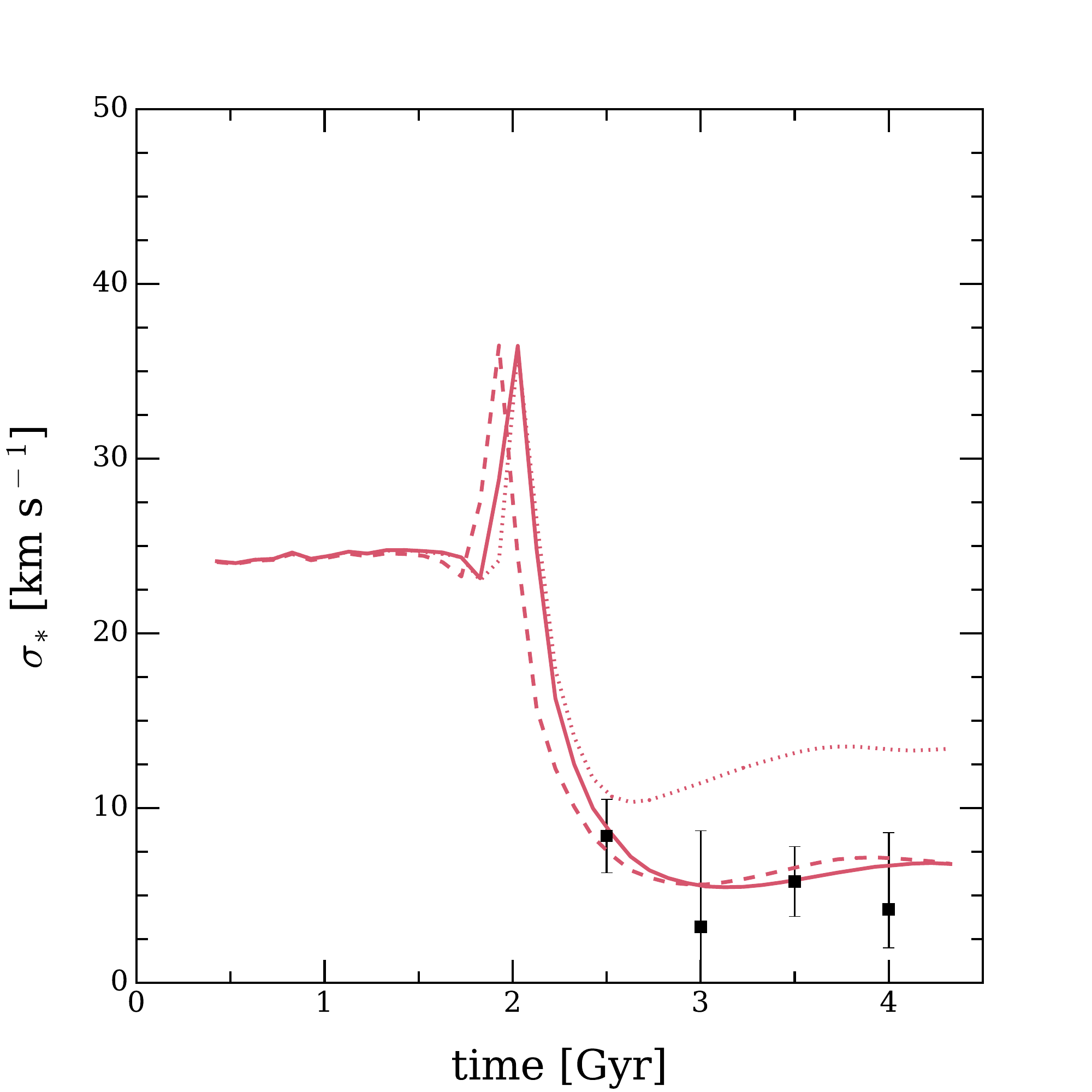}
\vspace{-.25cm}
\caption{Evolution of the stellar velocity dispersion with time at a fixed orbit (top panel) or at a fixed galaxy (bottom panel). In the top panel, the evolution is shown for 
G1, G2, and G3 (same color coding as previous plots) on the O2 orbit.
The bottom panel shows the evolution of the G2 galaxy on the O1 (dashed), 
O2 (solid), and O3 (dotted) orbits. Observational data are shown by 
black points with errorbars.}
\label{fig:O2G2}
\end{figure}

\begin{figure}
\includegraphics[width=0.47\textwidth]{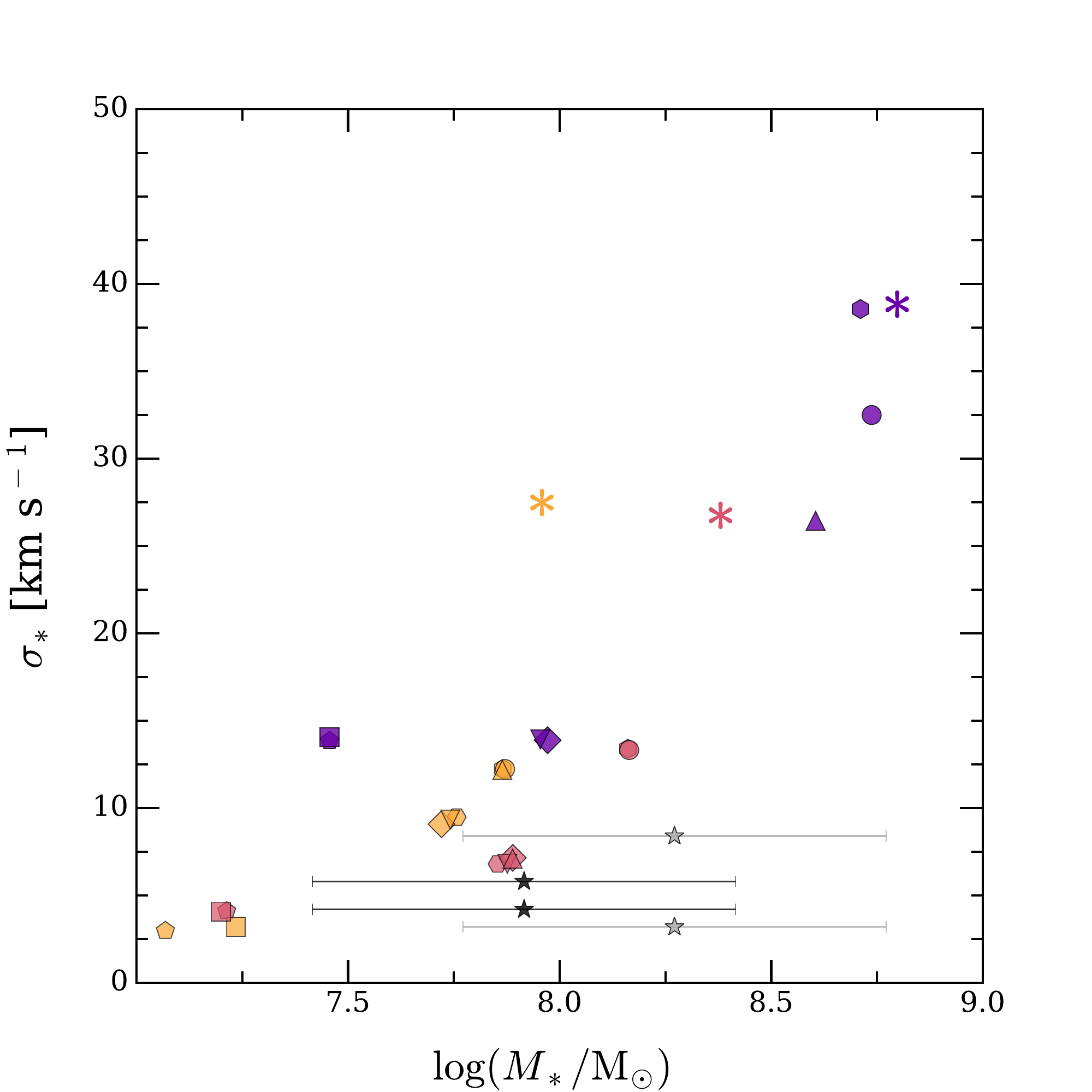}
\vspace{-.25cm}
\caption{Relation between stellar mass and stellar velocity dispersion for twenty four simulations (8 orbits for 3 galaxies).
Different symbols correspond to different orbits (see Table \ref{tab:orbits}), while color scheme is the same as in the previous plots. The asterisk represent the value of the velocity dispersion at infall. 
Grey and black stars represent observed data for DF2 and DF4, respectively (see text for more details).}
\label{fig:allorbits}
\end{figure}

Fig.~\ref{fig:O2G2} presents the evolution of the stellar velocity dispersion
either at a fixed orbit for the three galaxies or at a fixed galaxy on our three test orbits.
More specifically, the top panel shows G1,G2 and G3 all on orbit O2, while the 
bottom panel shows galaxy G2 on orbits O1 (dashed line), O2 (solid line), and O3 (dotted line).
Finally, Fig.~\ref{fig:allorbits} shows the results for the eight orbits (see Table \ref{tab:orbits}) on the three galaxies;
different symbols correspond to different orbits, while the color scheme is the same as in the previous plots. 

As the figures make clear, no matter the orbit, G3 always has a final velocity dispersion above 14~\kms, while G1 and G2 can reach dispersions well below 10~\kms on several orbits. The properties of the galaxy are indeed more important than the parameters of the orbit in setting the evolution of stellar velocity dispersion.

\begin{figure}
\includegraphics[width=0.47\textwidth]{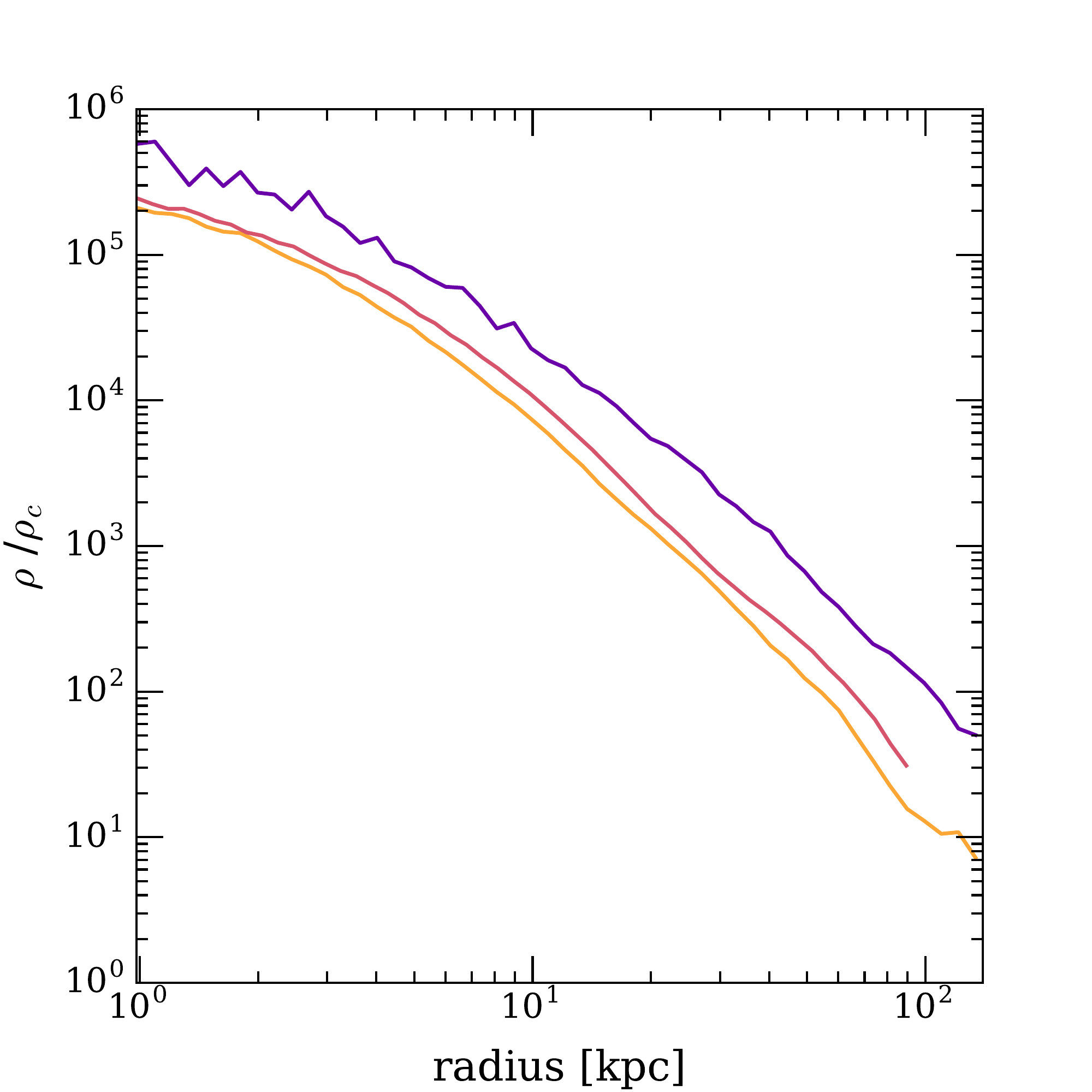}
\vspace{-.25cm}
\caption{Dark matter radial density profiles for G1, G2, and G3. 
The density is in units of $\rho_{\rm c}$ the critical density of the Universe.
The slopes of the density profiles at 2 per~cent of the virial radius ($\approx 1.3$
kpc) are 0.33, 0.64, 0.72 for the three galaxies, respectively \citep[see][for more details]{Maccio2020}.}
\label{fig:profiles}
\end{figure}

This is due to the more concentrated and partially ``cuspier'' dark matter density profile of G3 with respect to the other two galaxies, as shown in Fig.~\ref{fig:profiles}. 
All three galaxies have a density profile shallower than
predicted by pure $N$-body simulations \citep[e.g.][]{Dutton2014}. 
For example, the slopes of the density profile between 1 and 2 per~cent
of the virial radius are 0.33, 0.64, and 0.72 for the three galaxies, respectively \citep{Maccio2020}, while an NFW profile \citep{Navarro1996} always yields slopes steeper than -1.
Moreover, at all scales G3 has three to four times the mass of G1 or G2. 
These two effects make G3 more resilient to tidal forces
\citep{Read2006,Penarrubia2008,Ogiya2018}. As a consequence, the mass loss is reduced in G3, which then retains a larger velocity dispersion. 

Finally, despite the substantial tidal perturbation experienced by the galaxies near the center of the potential and once observed at a distance of hundred kpc from the center (similar to DF2 and DF4), we stress that the simulated galaxies do not show any clear evidence of tidal effects (e.g. stellar streams, flares, etc.). They present rather smooth morphologies in agreement with the observed ones for DF2 and DF4.

\section{Discussion and Conclusions}
\label{sec:conclusions}

In two recent papers, van Dokkum and collaborators have reported
the observation of two low-mass galaxies ($M_*\sim 10^8$~\Msun) with an extremely low stellar central velocity dispersion ($\sigma_*<10$~\kms).
This dispersion, when translated into a mass estimate and compared with the stellar mass within a radius of 8~kpc, seems to suggest a very low central dark matter content, almost consistent with no dark matter at all.
This contrasts with predictions from simulations and abundance
matching techniques \citep[e.g.][]{Moster2013} that predict galaxies
in this $M_*$ range to live in much more massive dark matter haloes with 
velocity dispersion close to 30~\kms and being strongly DM-dominated
in their central regions (up to 30-40 times more dark than visible matter).
These observations bring into question whether these DM-deficient galaxies can actually form
in a DM-dominated universe \citep[as implied by CMB observations][]{Planck2014}, or if they somehow challenge our current picture of galaxy and structure formation.

In this paper, we have used a combination of cosmological hydrodynamical simulations from the NIHAO project \citep{Wang2015} and direct interaction simulations \citep{Frings2017,Ogiya2018} to investigate whether tidal effects can explain the
dynamical properties of DF2 and DF4 within the confines of the classical galaxy formation scenario \citep[e.g.][]{Somerville2015}.

By comparing the observed velocities with the ones extracted from NIHAO galaxies, we have shown that objects with a $\sigma_*$ as low as the ones in DF2 and DF4 cannot be reproduced by our current simulations of field galaxies (i.e. central objects), despite NIHAO having quite a strong `halo expansion' as an effect of galaxy formation \citep{Tollet2016,Maccio2020}.
Since both DF2 and DF4 are satellite galaxies of NGC 1052, a massive elliptical galaxy, there have been suggestions that tidal effects could reduce the central DM density \citep{Ogiya2018,Jing2019}.

In our paper, for the first time, we have combined high-resolution cosmological simulations (with more than $10^5$ elements to describe DM, gas and stars) with
direct interaction simulations to study the effect of tidal forces on the DM distribution in DF2 and DF4 analogues. 

By injecting cosmological galaxies on orbits with a strong radial component, we have created galaxies capable of reproducing all the key properties of DF2 and DF4.
We have obtained a simulated central stellar velocity dispersion as low as
$5 \pm 2$~\kms, 
where the uncertainty arises from the random sampling of ~10 globular cluster analogues for computing the velocity dispersion.
Despite strong tidal effects near the center, simulated galaxies
do not show evidence of tidal perturbations (e.g. stellar streams) once observed at a distance of few hundreds kpc from the center, in agreement with the position and morphology of DF2 and DF4. 

The change in the stellar velocity dispersion from initial (i.e. in isolation) values around 30~\kms to the final ones of few \kms is related to the strong preferential removal of dark matter (with respect to stars) even from an inner region of 8~kpc in radius. The dark matter content is reduced by a factor of 20 to 100, while the stellar mass only changes by a factor of a few. 
As a result, the central DM to stellar mass ratio changes from 25-40 to values closer to unity.
We ascribe this differential change to the different orbits of the central stellar and dark matter particles, with the latter being in more radially elongated trajectories making them more susceptible to tidal forces \citep[e.g.][]{Jesseit2004,Penarrubia2010}.

Overall, our study shows how it is possible to create, in a dense environment, galaxies with extremely low central velocity dispersions, with stellar masses and sizes consistent with those of DF2 and DF4.
Galaxies lacking dark matter in a group environment may not represent a departure from current cosmological models. Further observational 
\citep{Montes2020}
and numerical investigations\footnote{A paper by \citet{Jackson2020} reaching similar conclusions was submitted on arXiv within seconds of our post. The calculation of the actual
probability of such an event is beyond the scope of this work.} must assess if the frequency of dark matter deficient galaxies is consistent with  typical satellite orbit distributions predicted by hierarchical structure formation.

\section*{Data availability}

The data underlying this article will be shared on reasonable request to the corresponding author.
\section*{Acknowledgements}

The authors  gratefully acknowledge the Gauss Centre for Supercomputing e.V. (www.gauss-centre.eu) 
for funding this project by providing computing time on the GCS Supercomputer SuperMUC at Leibniz
Supercomputing Centre (www.lrz.de) and the High Performance Computing resources at New York University Abu Dhabi.
AVM thanks J. Frings for sharing the software to set up the interaction simulations.
TB acknowledges support from the European Research Council under ERC-CoG grant CRAGSMAN-646955. SC acknowledges funding from the Natural Science and Engineering Research Council of Canada via their Discovery Grant program.

\newpage 



\bibliographystyle{mnras}
\bibliography{ref} 






\bsp	
\label{lastpage}
\end{document}